\documentstyle[sprocl, epsfig]{article}

\def\Journal#1#2#3#4{{#1} {\bf #2}, #3 (#4)}

\def\NIMA{{\em Nucl. Instrum. Methods} A}

\def\etal{{\it et\ al.}}

\def\ra{\rightarrow}

\def\be{\begin{equation}}
\def\ee{\end{equation}}
\def\bea{\begin{eqnarray}}
\def\eea{\end{eqnarray}}

\begin{document}

\title{THE ICARUS 50 l LAr TPC IN THE CERN $\nu$ BEAM}

\author{F.~ARNEODO, F.~CAVANNA, I.~DE~MITRI, D.~MAZZA, S.~PARLATI, 
S.~PETRERA, G.~PIANO~MORTARI, M.~VERDECCHIA}
\address{Dipartimento d Fisica e INFN, Universita' dell'Aquila, 
via Vetoio, L'Aquila, Italy},

\author{P.~BENETTI, A.~BORIO DI TIGLIOLE, E.~CALLIGARICH, E.~CESANA, 
R.~DOLFINI, 
F.~MAURI, C.~MONTANARI, A.~RAPPOLDI, G.L.~RASELLI, M.~TERRANI, P.~TORRE,
C.~VIGNOLI}
\address{Dipartimento di Fisica e INFN, Universita' di Pavia, via Bassi 6, 
Pavia, Italy}

\author{A.~BETTINI, C.~CARPANESE, S.~CENTRO, A.~PEPATO, F.~PIETROPAOLO, 
S.~VENTURA}
\address{Dipartimento di Fisica e INFN, Universita' di Padova, 
via Marzolo 8, Padova, Italy}

\author{M.~BONESINI, B.~BOSCHETTI, M.~CALVI, A.~CURIONI, D.~CAVALLI,
A.~FERRARI, P.~FERRARI, 
P.~NEGRI, M.~PAGANONI, A.~PULLIA, P.~SALA, S.~RAGAZZI, N.~REDAELLI,
T.~TABARELLI~DE~FATIS, F.~TERRANOVA, A.~TONAZZO}
\address{Dipartimento di Fisica e 
INFN, Universita' di Milano, via Celoria 16, Milano, Italy}

\author{F.~CASAGRANDE}
\address{Laboratori Nazionali dell'INFN di Frascati, via Fermi 40, 
Frascati (RM), Italy}

\author{P.~CENNINI, A.~PUCCINI, A.~RUBBIA, C.~RUBBIA, J.P.~REVOL}
\address{CERN, CH-1211 Geneva 23, Switzerland}

\author{D.~CLINE, C.~MATTHEY, S.~OTWINOWSKI, J.~PARK, H.~WANG, J.~WOO}
\address{Department of Physics, UCLA, Los Angeles, CA 90024, USA}

\author{J.~GIVOLETTI}
\address{Dipartimento di Ingegneria, Universita' di Siena, via Roma 56,
Siena, Italy}

\author{A.~LAMARINA, R.~PERIALE, P.~PICCHI}
\address{Dipartimento di Fisica e INFN, Universita' di Torino,
 via Giuria 1, Torino, Italy}

\author{G.~MANNOCCHI, L.~PERIALE, S.~SUZUKI}
\address{ICGF-CNR, Corso Fiume 4, Torino, Italy}

\author{F.~SERGIAMPIETRI}
\address{INFN Pisa, via Livornese 1291, San Piero 
a Grado (PI), Italy}

\maketitle\abstracts{
The 50 liter liquid Argon TPC is a detector built and  
successfully operated at CERN for R\&D purposes within the ICARUS programme. 
In the year 1997 it has been exposed at the CERN neutrino beam for the entire
SPS neutrino run period as proposed and approved at the SPSLC of January 
1997 \cite{fiftyliter}. The detector, complemented with scintillators acting
as veto, trigger counters and pre-shower counters, was 
installed in front of the NOMAD detector.
The year 1997 was scheduled to be the last for the operation of the West Area 
Neutrino Facility. It was important to take this last opportunity 
for a parasitic exposure, which did not interfere with running experiments, 
of an already existing and operating liquid Argon TPC. 
As we had expected, 
the collected data brought important information for a better understanding 
of the performance of liquid Argon TPC's which should be useful for the entire 
ICARUS program.
}

\section{Physics motivations}\label{sec:phys}

An important part of the ICARUS physics program aims at the measure 
of neutrino oscillations through the detection of LBL neutrino 
interactions\cite{icalbl}. Both the $\nu_{\mu} \ra \nu_e$ and the 
$\nu_{\mu} \ra \nu_{\tau}$  channels have been condidered.

The possibility to isolate $\nu_{\tau}$ charged current interactions 
by means of kinematic cuts
on the reconstructed events has been studied in details. It turned out that
the main limitation of this method comes from the bad knowledge of the effect
of Fermi motion and nuclear rescattering on the final state 
kinematics\cite{ferflu}.
It was therefore very important to acquire experimental data in order
to tune the Monte Carlo models.

The exposure of a small Lar TPC prototype at the CERN neutrino beam with the
collection of a substantial sample of quasi-elastic 
$\nu_{\mu} + n \ra p + \mu^-$ events matched the above requirements.
The main physics items which we planned to study in this test 
were the following:
\begin{itemize}
\item appearance of nuclear fragments at the vertex of the
interaction in events having the $\mu-p$ topology;
\item measurement of the $\nu-\mu-p$ acoplanarity and missing transverse 
momentum in events with the $\mu-p$ topology interactions, in order 
to assess Fermi motion and proton re-scattering inside the nucleus;
\item a preliminary of evaluation of $e/\gamma$ and $e/\pi^0$ separation 
capability by means of the measurement of the specific 
ionisation on the wires, however, limited by the size of the chamber.
\end{itemize}

A further goal of this test was to gain experience with real
neutrino events. This experience has provided general information useful 
for the study of atmospheric neutrinos, proton decay and high energy neutrinos
from CERN. For example the test has led to the optimization 
of the readout chain in view of best extracting the features of these events.

\section{Setup of the test}

\subsection{The ICARUS 50 liter liquid Argon TPC}

The detector structure consists in a stainless steel cylindrical main vessel, 
70 cm diameter, 90 cm height, whose upper face is an UHV flange housing the 
feed-through's  for vacuum, liquid Ar filling, high voltages and read-out 
electronics. Inside the main vessel an ICARUS type liquid argon TPC is mounted.
The TPC has the shape of a 
parallelepiped  whose opposite horizontal faces ($32\times 32\rm\ cm^2$) 
act as cathode and anode, while the side faces, 47~cm  long, support the 
field-shaping electrodes. The mass of the liquid argon contained in the active 
volume is 65 kg. 
\par
The read-out electrodes, forming the anode, are two parallel wire planes 
spaced by 4~mm. Each plane is made of stainless steel wires, 
100~$\mu\rm m$ diameter and 2.54~mm pitch. The first 
plane (facing the drift volume) works in induction mode while the second 
collects the drifting electrons.
The wire direction on the induction plane runs orthogonally to that on the 
collection plane (cfr. with the ICARUS 600 ton set-up where the three planes 
are foreseen at $60^o$ to each other).
The wire geometry is the simplest version of the ICARUS readout 
technique \cite{icardout} since both the screening grid and field wires in 
between sense wires have been eliminated.
The wires are soldered on a vetronite frame which supports also the high 
voltage distribution and the de-coupling capacitors. 
\par
The front-end electronics for the 256 read-out channels is mounted directly on 
the frame in order to minimise the input capacitance of the pre-amplifiers 
which are foreseen to work immersed in liquid Argon \cite{larfend}.

The cathode and the field shaping electrodes have been obtained with a printed 
board technique on a vetronite support. The printed boards are glued on a 
honeycomb structure which ensures their rigidity. The field-shaping electrodes 
are horizontal strips, 1.27~cm wide, spaced by 
2.54~cm. A high-voltage divider, made by a series of resistors 
($100\rm\ M\Omega$ each) interconnecting the strips, supplies the correct 
potential to each strip. The drift high voltage ($\leq 15\rm\ kV$) 
is brought to the cathode through a commercial ceramic feed-through.

The whole main vessel is immersed in a thermal bath of commercial liquid Ar, 
contained in an open air stainless steel dewar. The argon evaporation from 
the open-air dewar is about 50 liters/day.
The detector is equipped with a standard ICARUS recirculation-purification 
system \cite{ricirc}. The full set-up was
already tested and successfully operated in CERN for R\&D purposes.

\subsection{Installation in the CERN neutrino beam}

\begin{figure}[htb]
\begin{center}
\epsfig{figure=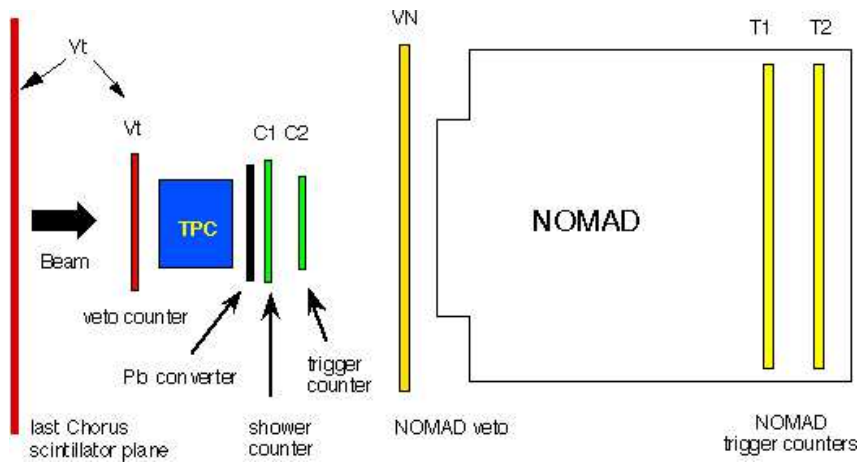,height=6cm}
\caption{Experimental set-up of the 50 liter Liquid Argon TPC 
in the CERN $\nu$ beam.}
\label{fig:layout}
\end{center}
\end{figure}

The layout of the experimental set-up is sketched in fig. \ref{fig:layout}. 
Since muon identification and momentum measurement were essential for 
the studies we wished to perform, the chamber had to be complemented with a 
muon identifier and a spectrometer. Together with the NOMAD collaboration 
we came to the conclusion that the NOMAD detector could perform these tasks.

Given the size of the TPC and the radiation length of liquid Ar 
($X_0=14\rm\ cm$), $\pi^0$'s 
produced in neutrino interactions may escape detection, thus $\mu-p-\pi^0$ 
events, with undetected $\pi^0$, could fake quasi-elastic interactions 
with large missing $P_T$. The 
installation of a pre-shower counter downstream of the TPC was implemented in 
order to attempt to identify gammas leaving the chamber.

The detector has been installed in CERN hall 191 between CHORUS and NOMAD 
detectors. Dewar, argon purification system, vacuum pumps, veto and trigger 
scintillators, trigger and read-out electronics and data acquisition system, 
have been placed on a  platform at 3.90 m from ground level. 
An argon tank of 5000 liters has been installed outside of the hall. Liquid 
argon, for periodic re-filling, has been brought from the outer tank to the 
dewar containing the TPC by isolated pipes running in protected position on 
the floor of the hall and reaching the platform along its scaffolding. 

Immediately upstream of the chamber a double plane of scintillators, acting as 
a veto counter, have been installed; downstream of the chamber there is a 6~mm 
thick lead sheet followed by a plane of scintillators, acting as trigger and 
pre-shower counter.

\subsection{Set-up of read-out and data aquisition}

A standard ICARUS read-out of the 256 channels was implemented allowing to
acquire up to 4 triggers per neutrino burst. Being the electron drift velocity 
about $1\rm\ mm/\mu s$ and the signals sampled by FADC at 400~ns, to span the 
total drift space one needs about 2 kb per channel or about 600~kb/event. 
\par
The events are written in raw-data format without any zero suppression;
they are stored locally on disk and automatically transfered to
the main CERN tape facility using the network.
The trigger of the TPC read-out was based on coincidences between the
down-stream scintillators and the NOMAD muon trigger planes. The upstream
scintillators, ORed with signals from the CHORUS muon spectrometer
scintillator planes were used to veto passing through particles (mostly
muons produced by neutrino interactions upstream). 
At he beginning of each spill NOMAD sent, as pulse trains on standard BNC 
cables, run number and burst number, which have been recorded by scalers 
read-out by the TPC acquisition system. This allowed off-line matching of 
events in the TPC with muons reconstructed by NOMAD.
For some specific period of times (calibration, gamma rays source, etc...) 
we also self-triggered the chamber exploiting the analog-sum of the 
collection wires.

The dead time of the TPC data acquisition was measured to be
lower than 5\%, while the NOMAD dead time is around 15\%.

\section{Results obtained so far with the test}

\subsection{Experience with the chamber operation}

\begin{figure}[htb]
\begin{center}
\epsfig{figure=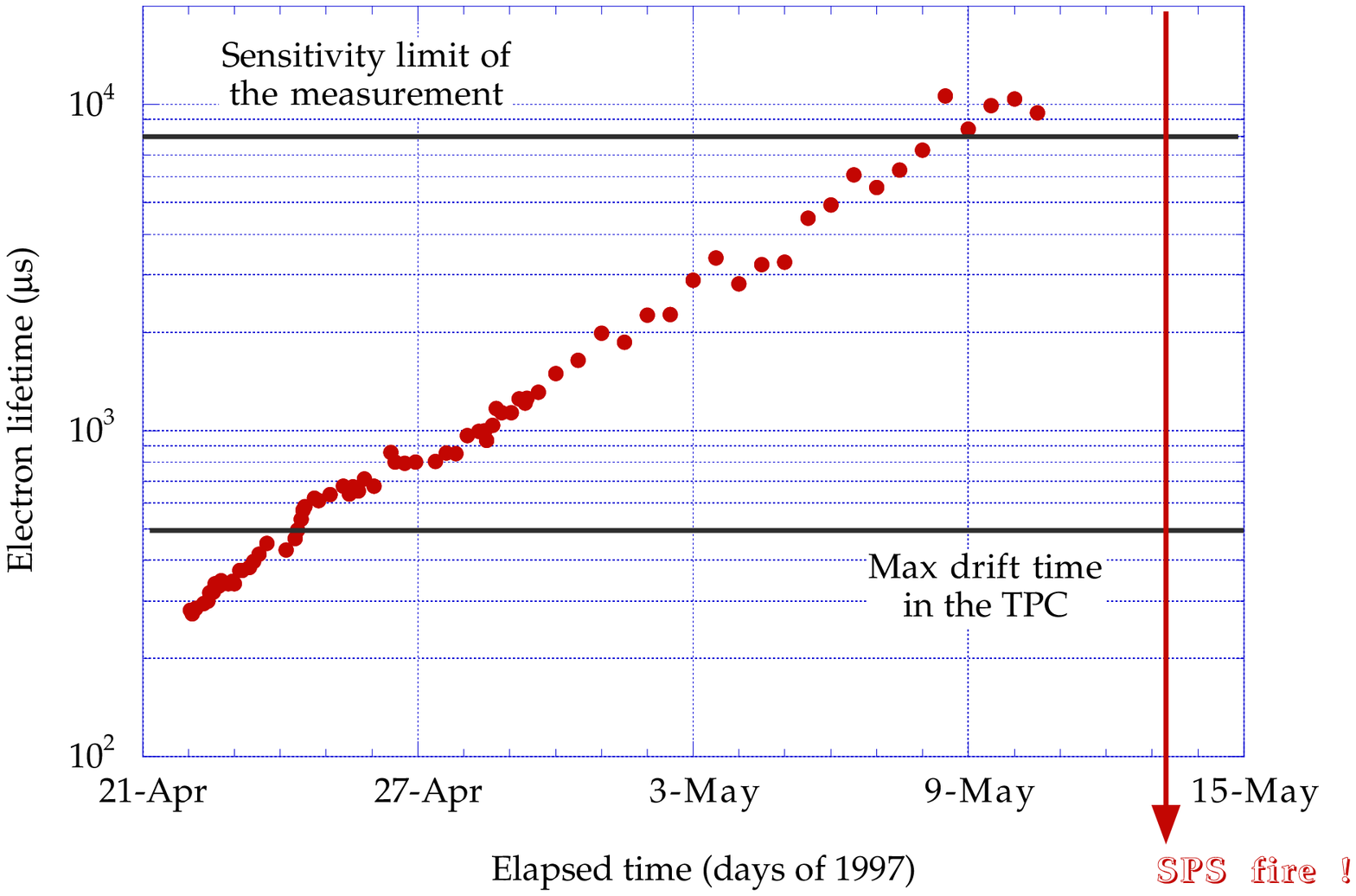,height=6cm}
\caption{Lifetime of the drifting electrons in the 50 liter 
Liquid Argon TPC at the CERN $\nu$ beam. The filling of the chamber with LAr 
was performed on the fourth of April 1997.}
\label{fig:lifetime}
\end{center}
\end{figure}

The detector was filled with ultra pure liquid argon flown 
through the ICARUS purification system in liquid phase 
(at a rate of 60 liter/hour) \cite{liquidphase}. 
\begin{itemize}
\item The initial lifetime was about 100 $\mu\rm s$.
The recirculation/purification system, circulating about 5 liters of LAr/hour 
allowed to increase this value to more than 8 ms in three weeks 
(see fig. \ref{fig:lifetime}). 
This performance was excellent since the
maximum drift time was 400 $\mu\rm s$ and therefore the attentuation of the
ionization over the drift distance was negligible.
\item This purity level was kept stable for all the running period of 9~months.
\item The same recirculation system allowed to dissolve in pure LAr a small 
concentration of TMG (few ppm) necessary to recover the linearity response 
in deposited energy \cite{tmg}.
\item The total LAr consumption necessary to circulate the pure LAr and to 
compensate the heat losses of the set-up was about 200 liter/day.
\item The planned operating value for the drift high voltage was 25 kV to 
reach the nominal drift field of 500 V/cm. Unfortunately we have experienced 
some instabilities of uncertain origin which forced us to operate at a safe 
value of 10 kV. We believe that most likely this limitation was due to 
residual humidity in the HV feed-through at the level of the LAr interface. 
In the newly designed feed-through this problem should not arise. Given the 
achieved argon purity and the relatively short drift length, the reduced 
drift field did not affect significantly the performance of the chamber. 
\item The new wire chamber configuration appeared to be sound. The visibility 
of the induction signal was optimized by varying the ratio of the fields
in the drift and gap regions.
\end{itemize}

\begin{figure}[htb]
\begin{center}
\epsfig{figure=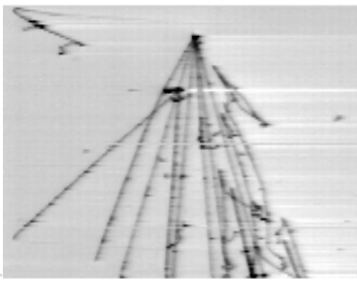,height=4cm,width=5cm}
\caption{An example of recorded neutrino interaction in the 50 liter 
Liquid Argon chamber prototype located at the CERN $\nu$ beam. The
neutrino comes from the top of the picture. The horizontal axis
is the time axis (drift direction) and vertically is the wire number.
The visible area corresponds to $47 \times 32$ $cm^2$}
\label{fig:icarusevent}
\end{center}
\end{figure}

\subsection{Experience gained with the front-end electronics}

In fig. \ref{fig:icarusevent} we show, as an example, a neutrino event 
recorded with the 50 liter LAr TPC exposed at the CERN neutrino beam.
From the inspection of many similar events and from the analysis of 
simulations of neutrino interactions in Liquid Argon, we decided to 
adopt a front-end configuration based on a current amplifier with a 
feed-back resistance of $R_f=5\rm M\Omega$. 

With this configuration we 
obtained a signal-to-noise ratio $S/N=11$ with mip signal equivalent to 
10 ADC counts. This choice was satisfactory because the risk of pile-up 
in events containing electromagnetic showers was highly suppressed. 
This is essentially due to the fact that the duration
of the signal was comparable with the distance between
tracks in electromagnetic showers. 

A further consequence of the front-end choice was that the digital dynamic 
range of 8 bits was sufficient even for the case of electromagnetic showers. 
A detailed analyis of the performance of the adopted front-end scheme can
be found in reference \cite{cennini}.

\subsection{Preliminary analysis of quasi-elastic neutrino events}

\begin{figure}[htb]
\begin{center}
\epsfig{figure=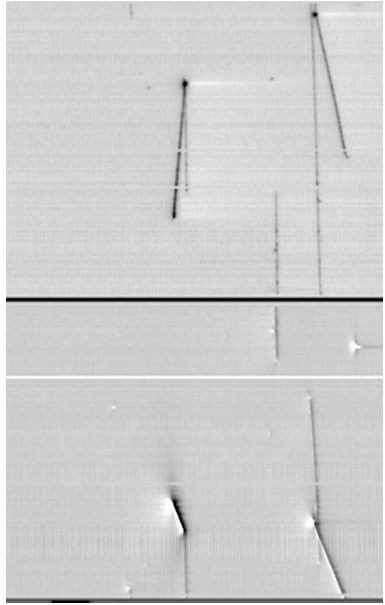,height=8cm,width=5cm}
\caption{An example of neutrino quasi-elastic interaction in the 50 liter 
Liquid Argon TPC. The two orthogonal views are shown.
The neutrino comes from the top of the picture. The horizontal axis
is the time axis (drift direction) and vertically is the wire number.
The visible areas corresponds to $47 \times 32$ $cm^2$}
\label{fig:qeev}
\end{center}
\end{figure}

During the 1997 CERN-SPS neutrino run we collected more than $10^5$ triggers.
The sample of all triggers accumulated during the run has been 
visually scanned to select neutrino interaction candidates. 
The whole scanning has been performed three times in order to ensure 
a high efficiency. About 9000 charged current $\nu_{\mu}$ events have been
identified in agreement with the expectation for the $1.2~10^{19}$ pot 
integrated in the 1997 run. The quasi-elastic candidates amount to 350.
A typical example of quasi-elastic candidate is shown in fig. \ref{fig:qeev}.

Analysis is presently being performed on a subsample of quasi-elastic
candidates which satisfy the following criteria:
\begin{itemize}
\item a vertex is identified with at most two tracks leaving it;
\item one of the track has to be a mip exiting the chamber; its has to 
be recognized as a muon by the NOMAD spectrometer;
\item the second track, if any, has to be recognized as a proton  $dE/dx$;
\item no other hadron or gamma has to be present.
\end{itemize}

The sample selected in this way amounts to about 150 clean quasi-elastic 
$\nu_\mu n\rightarrow \mu^-p$ events and is expected to contain about 1/2 of 
the total quasi-elastic events and some contamination deu to unseen $\pi^0$.
This is probably a biased sample of quasi-elastics, however it has the 
advantage of consisting of simple events in which particles are well
identified and measured. 

\begin{figure}[htb]
\begin{center}
\epsfig{figure=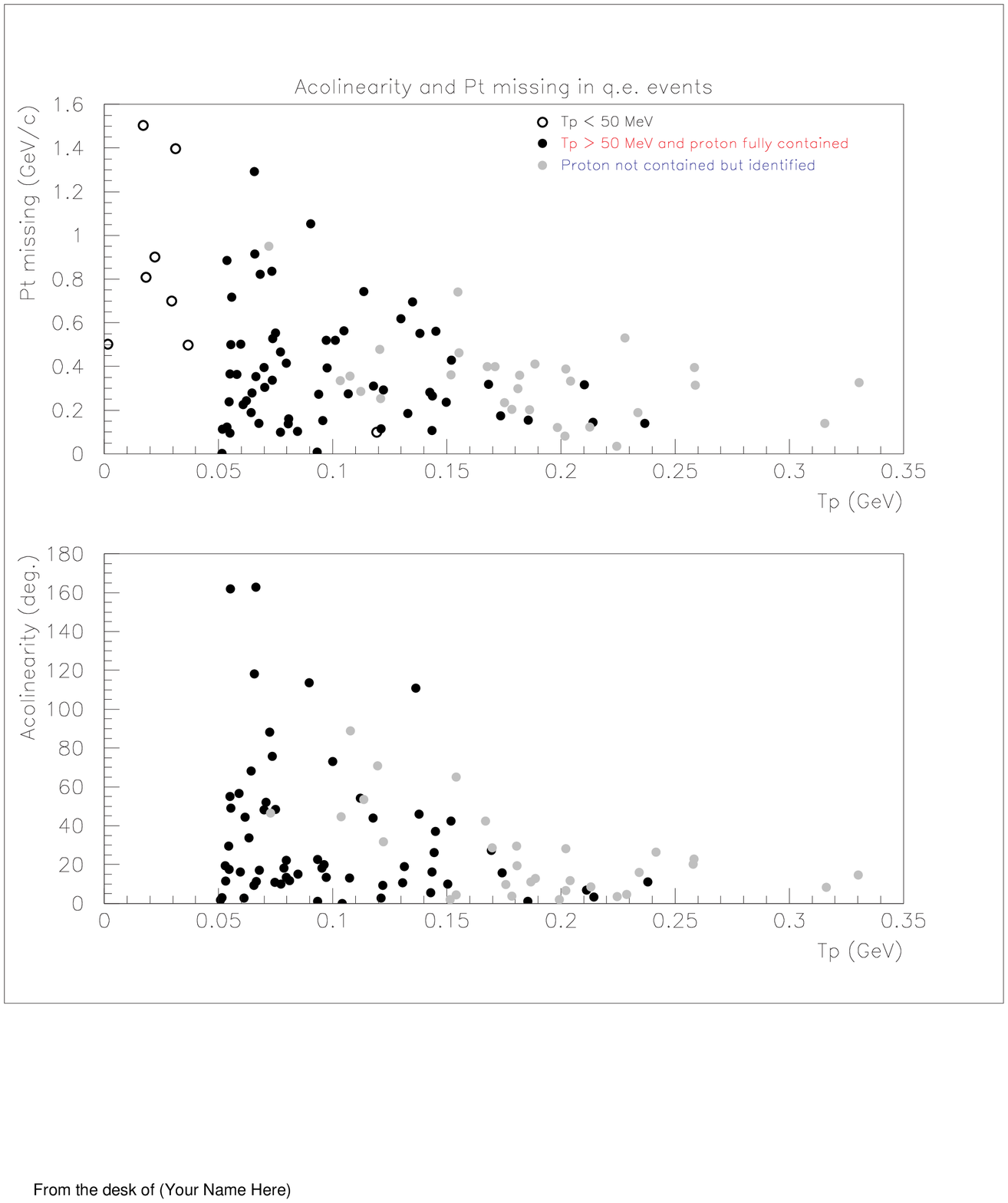,height=9cm}
\caption{Missing $p_t$ and acolinearity distributions as a function of the
reconstructed kinetic energy of the leading proton.}
\label{fig:ptmiss}
\end{center}
\end{figure}

The preliminary results from the analysis of these events are summarized in 
fig. \ref{fig:ptmiss}; it shows the reconstructed momentum unbalance of
the interaction in the transverse plane. Only a fraction of the events with 
small proton kinetic energy present a large missing $p_t$ and acolinearity.

At proton kinetic energy larger then 150 MeV all the analized events
present a good behaviour, namely acolinearity and missing $p_t$
compatible with the Fermi motion without hard rescattering. A small part 
of the large missing $p_t$ sample could be due to background.

A detailed simulation of these events, based on the FLUKA 97.5 pakage,
is underway to determine the reliability of our Monte Carlo models. 
Preliminary comparisons are satisfactory.

\section*{Acknowledgments}

The authors wish to thanks the organizers of the conference for having given 
the opportunity to present our work among some of the most exciting 
developments in the field of new detectors.

A special thank goes to the NOMAD collaboration for their unlimited support 
during the setting-up of the test, the data-taking and the event analysis.

\end{document}